\documentclass[aps,twocolumn,showpacs,fleqn]{revtex4}

\usepackage{amsmath,amssymb}
\usepackage{graphicx}

\renewcommand{\vec}{\mathbf}

\def\fig#1{Fig.\,\ref{#1}}

\def\Eq#1{Equation~(\ref{#1})}
\def\eq#1{Eq.\,(\ref{#1})}

\def\fig#1{Fig.\,\ref{#1}}

\def\vz{\vec{z}}\def\vf{\vec{f}}
\def\vrt{\vec{r}}\def\vrs{\vec{r}'}\def\rt{r}\def\rs{r'}

\begin{document}

\title{Spatial correlations in finite samples revealed by Coulomb explosion}
\author{Ulf Saalmann}\author{Alexey Mikaberidze\footnote{current address: Institute of Integrative Biology, ETH Zurich, Universit\"atstr.\,16, 8092 Zurich, Switzerland}}\author{Jan M. Rost}
\affiliation{Max Planck Institute for the Physics of Complex Systems\\
 N\"othnitzer Stra{\ss}e 38, 01187 Dresden, Germany }

\begin{abstract}\noindent
We demonstrate that fast removal of many electrons uncovers initial correlations of atoms in a finite sample through a pronounced peak in the kinetic-energy spectrum of the exploding ions. This maximum is the result of an intricate interplay between the composition of the system from discrete particles and its boundary. 
The formation of the peak can be described analytically, allowing for correlations beyond a mean-field reference model.
It can be experimentally detected with short and intense light pulses from 4th-generation light sources. 
\end{abstract}

\maketitle\noindent
Correlation effects beyond the mean-field description are important in understanding the detailed structure of matter. A major scheme has been the ``exchange-correlation hole'' indispensable in electronic-structure theory \cite{fu95}. While this research focuses on bound, mostly even on the ground state, very intense light pulses now available for XUV to X-ray energies \cite{mcth10}, create situations to probe correlation in the continuum. One phenomenon in this context is massively parallel ionization \cite{gnsa+12} of a cluster or large molecule: Many electrons are ionized almost simultaneously to an energy high enough such that the attractive ionic background does not play a role on the short time scale of energy exchange and correlation of electrons in the continuum.

Complementarily, one may ask how the many ions behave whose dynamics develops on a longer time scale when most of the electrons have left.
Do the ions also develop a correlated motion in the continuum\,---\,and if so, what is its characteristics in terms of observables, e.g., the energy spectrum of the ions? The simplest description well-known from plasma physics approximates these ions as a continuous charge density. The corresponding spectrum provides a comparison for highlighting the differences to a correlated continuum calculation. They occur for two reasons: Firstly, the granularity of our system, i.e., the fact that we deal with real particles which typically have a minimal separation $\delta$ and therefore give rise to a  classical ``correlation hole''.
Moreover, the ionic system is finite (for simplicity we assume a spherical shape) with an edge. The softness of this edge, i.e., the distance $a$ over which the charge density decreases to zero, is the second spatial scale which influences the ionic correlations in the continuum. In fact, we will show that the ion dynamics exhibits a crossover, characterized by the ratio $a/\delta$, from a dominant influence of the correlation hole if $a\,{\ll}\,\delta$ (hard edge) to the dominant influence of the edge for $a\,{\gg}\,\delta$ (soft edge).

\begin{figure}[b!]
\centerline{\includegraphics[width=0.8\columnwidth]{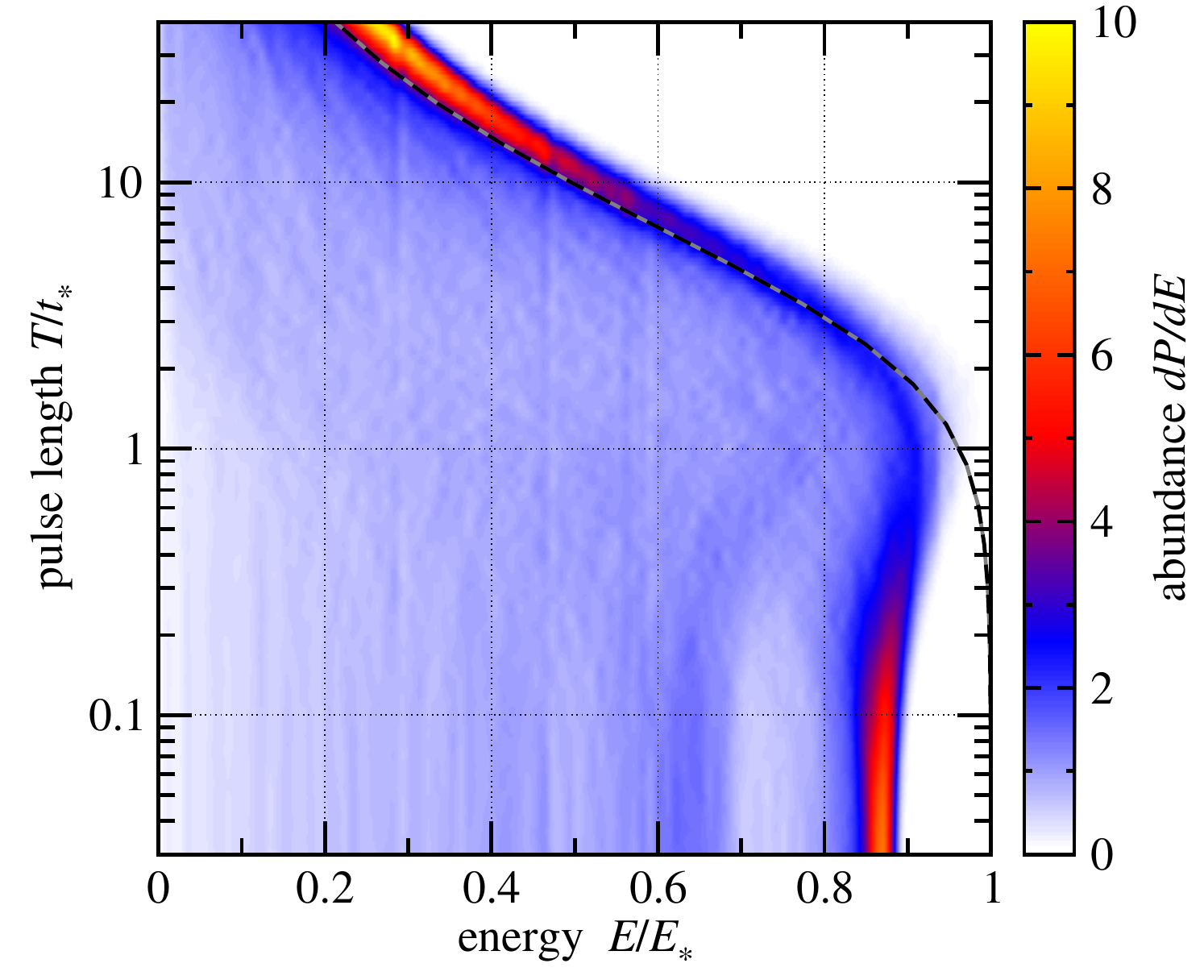}}
 \caption{(Color online) Energy spectra as obtained from an exploding LJ cluster with 1000 particles
 for various pulse durations $T$ given in units of the expansion time $t_{\star}$, cf.\ \eq{eq:ticm}. Energy is given in terms of $E_{\star}$ the highest energy for homogeneously charged sphere, cf.\ \eq{eq:edcm2}.
 The dotted line marks the surface energy of a sphere, which is charged homogeneously according to a Gaussian pulse (see text).}
\label{fig:spec-lj1000-pulse}
\end{figure}%
Surprisingly, this crossover can be probed by simply varying the pulse length of the ionizing light pulse in the correct parameter regime: A long pulse (for reasonable target systems this would be of the order of a few hundred femtoseconds) will slowly ionize the cluster and give the ions created near the surface time to move outward.
 Thereby, a soft edge in terms of decreasing density of ions is created. On the other hand, a short violent pulse
(of a few femtoseconds only) will start the motion of all ions created at once leading to correlation-hole dominated dynamics.
 
We set the stage by defining the mean-field reference system, a homogeneously charged sphere of charge $Q$ and mass $M$. It undergoes Coulomb explosion, doubling its initial radius $R$ in time
\begin{equation}\label{eq:ticm}
t_{\star}=\left(1{+}\ln(\sqrt{2}{+}1)\big/\sqrt{2}\right)\sqrt{MR^{3}/Q^{2}}\,.
\end{equation}
Making use of Gauss' law and the fact that shells of charge in a homogeneously charged sphere do not overtake each other, one can directly map the initial potential energy due to the enclosed charge at a radius $\rt$ to the final kinetic energy $E$ per unit charge $E(\rt) = Q \rt^{2}/R^{3}$. Evaluating the standard expression for the energy distribution 
\begin{equation}\label{eq:spec}
\frac{dP}{dE}=\frac{3}{R^{3}}\int_{0}^{R}\!{\rm d}\rt\,\rt^{2}\,\delta\big(E-E(\rt)\big)
\end{equation}
yields \cite{niam+01,kadu+03,issa+06}
\begin{subequations}\label{eq:edcm}\begin{align}\label{eq:edcm1}
\frac{dP}{dE} & =
\left\{\begin{array}{ccc}
(3/2)\sqrt{E/E_{\star}{\!}^{3}} & \mbox{for} & E\,{<}\,E_{\star} \\
0 & \mbox{for} & E\,{>}\,E_{\star}
\end{array}\right.
\\ \label{eq:edcm2} E_{\star}&=Q/R.
\end{align}\end{subequations}
In terms of these scales for energy $E_{\star}$ and time $t_{\star}$, we show in \fig{fig:spec-lj1000-pulse} kinetic-energy spectra of Lennard-Jones (LJ) clusters \footnote{LJ clusters imply that the ground state configuration of those clusters is obtained by minimizing the LJ interaction between the atoms.} with 1000 particles for various effective pulse duration $T$. Hereby, effective pulse duration refers to the time in which the particles get charged. 
We select randomly a charging time for each particle in such a way that the ensemble averaged charge grows as
$\langle Q_{t}\rangle=Q[1{+}{\rm erf}(t/T)]/2$ which, e.\,g., corresponds to ionization of atoms in a cluster by single-photon ionization, as it would occur for a sample exposed to intense free-electron laser radiation in a Gaussian pulse \cite{gnsa+11}.

\begin{figure}[t!]
 \centerline{\includegraphics[width=\columnwidth]{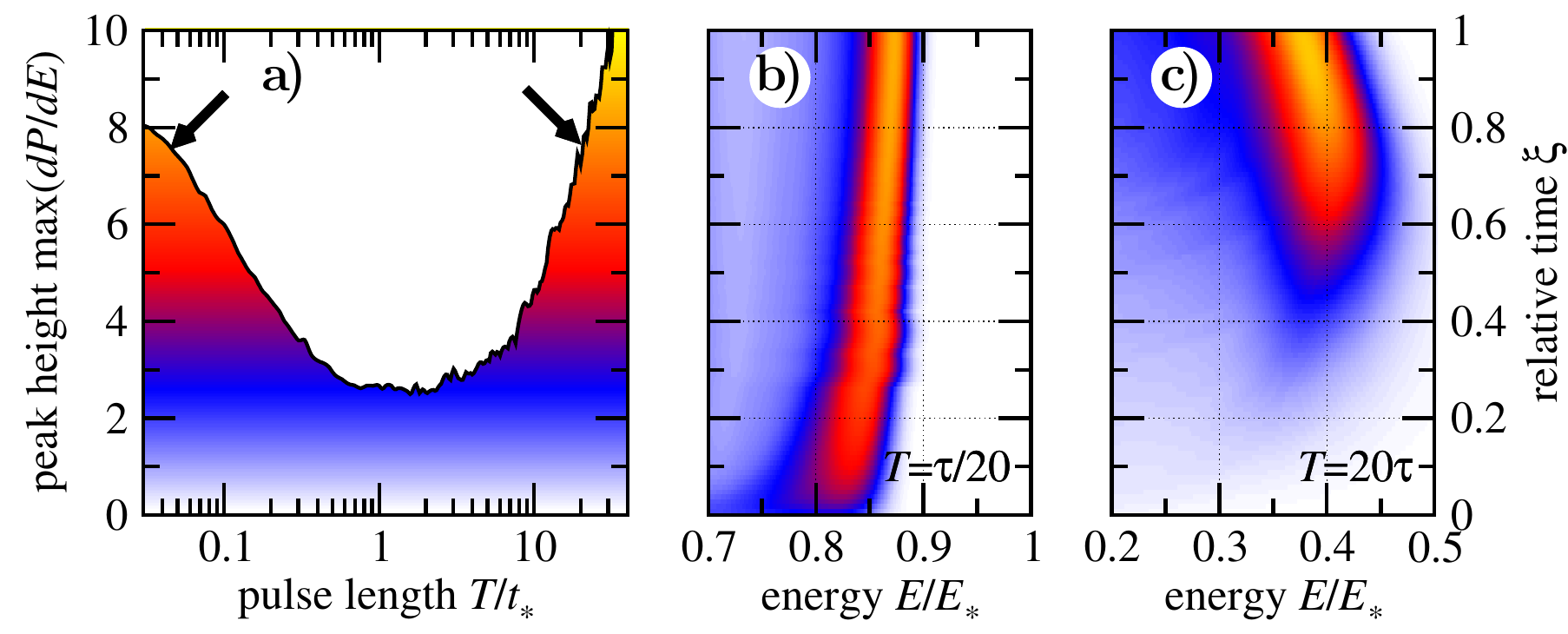}}
 \caption{(Color online) The maximal kinetic energy of the ions in Fig.\,\ref{fig:spec-lj1000-pulse} for different pulse lengths $T/t_{\star}$ (a). The two arrows indicate the pulse lengths $T = t_{\star}/20$ and $T = 20t_{\star}$ for which the evolution of the spectra $dP/dE$ are shown in (b) and (c) respectively as a function
 of energy and time $\xi\in [0,1]$. The time variable $\xi = E(t)/E(\infty)$ describes how far the Coulomb explosion has evolved in terms of the kinetic energy $E(t)$ of an expanding homogeneously charged sphere.}
\label{fig:spec-time}
\end{figure}%
One clearly sees in \fig{fig:spec-lj1000-pulse} a cross-over behavior with two ``hot spots'' at short and long pulses (cf.\ also \fig{fig:spec-time}a which shows the height of the ridge as a function of pulse length $T$), always close to the maximal energy possible at a given pulse length $T$. The \emph{long-pulse\/} regime is compatible with a mean-field dominated dynamics and has been described previously: This \emph{mean-field peak} in the kinetic-energy spectrum was observed for spherical objects with a washed-out edge, i.\,e., a non-uniform radial density distribution. It allows faster inner regions of exploding charge density to overtake slower outer ones, 
leading to radial caustics, as revealed by means of a kinetic model \cite{koby05byko05}. Under suitable conditions this caustic can even lead to a shock shell \cite{kadu+03}.
Pre-exploding the sample \cite{pefo+05} softens the edge of the system to  a gradually decreasing density, which is indeed close to the situation for long pulses shown in \fig{fig:spec-lj1000-pulse}.
 Charging at low densities as effected by a long laser pulse reduces the energy deposition in the sample and leads to a global decrease of 
 kinetic energies for long pulses, also visible in the spectrum for a homogeneously charged sphere (dashed line in \fig{fig:spec-lj1000-pulse}).

\begin{figure}[b!]
 \centerline{\includegraphics[width=\columnwidth]{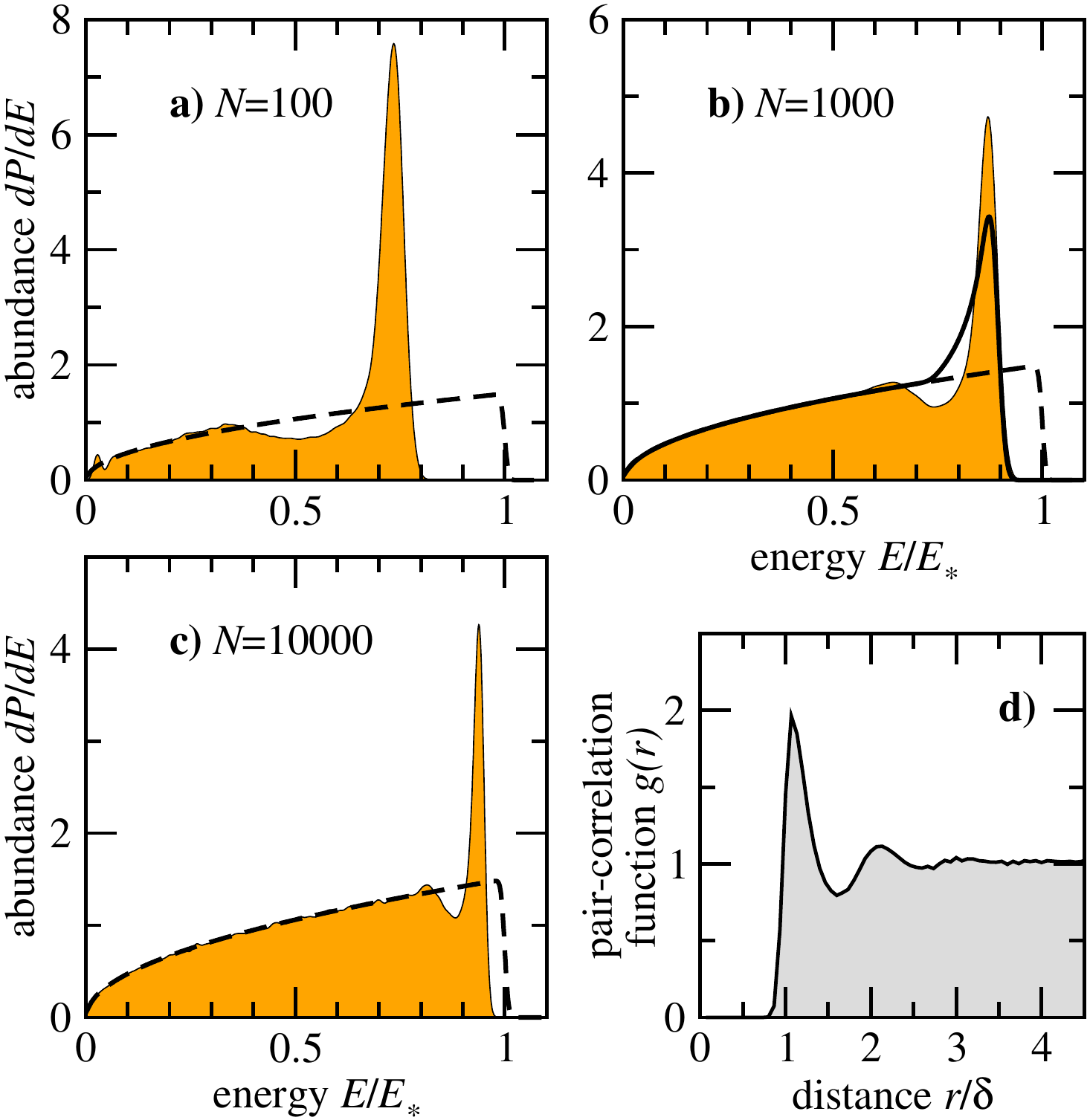}}
 \caption{(Color online) Spectra  from numerical propagation of systems with $N=100$, 1000 and 10000 ions, respectively, are shown by orange/gray-shaded areas (a--c).
The mean-field model as given by Eq.\,\eqref{eq:edcm} is shown with dashed lines. 
Considering a correlation hole in the mean-field description by using \eq{eq:emfch} in \eq{eq:spec} yields the spectrum shown as solid line (b).
Additionally the pair-correlation function $g(r)$ according to \eq{eq:mpcf} is shown for the cluster with $N{=}1000$ individual ions (d).}
\label{fig:spec-100-1000-10000}
\end{figure}%
What has not been seen before is the sharp maximum for \emph{short pulses\/} whose origin is fundamentally different from the mean-field peak at long pulses. We will refer to this maximum as the \emph{granularity peak} since its origin is the granularity of the system in connection with the fact that it is finite and has an edge. Indeed both features, the mean-field and the granularity peak are surface phenomena. This fact can be easily read off from \fig{fig:spec-lj1000-pulse}, since for the respective pulse length the peaks are located near the maximal energy contributed by surface ions. 
This is further confirmed by comparing the mean-field spectra according to \eq{eq:edcm} and the numerical ones from individual ions \footnote{The ions have been placed radially on equal-volume shells guaranteeing that the resulting granular density approximates the homogeneous density as close as possible. The angular positions of the ions are relaxed to minimize their interaction energy given by mutual LJ potentials.} in \fig{fig:spec-100-1000-10000}a-c. Formally, these spectra correspond to cuts at $T = 0$ in \fig{fig:spec-lj1000-pulse}, but for different cluster sizes, namely 100, 1000 and 10000 atoms, respectively. One sees, that the spectrum deviates less from the mean-field prediction for larger cluster size and correspondingly smaller surface-to-volume ratio, nevertheless the granularity peak remains clearly visible.

That mean-field and granularity peak are of different origin can be seen, e.g., from the formation of the maxima in time: 
The mean-field peak builds up gradually in time and appears towards the end of the explosion (\fig{fig:spec-time}c) since the inner density needs time to overtake the outer one. The granularity peak in case of a short pulse, on the other hand, appears right from the beginning (\fig{fig:spec-time}b) suggesting that it can be understood from the forces acting initially on the individual ions. This is indeed the case as we will see.

\begin{figure}[b!]
 \centerline{\includegraphics[width=0.75\columnwidth]{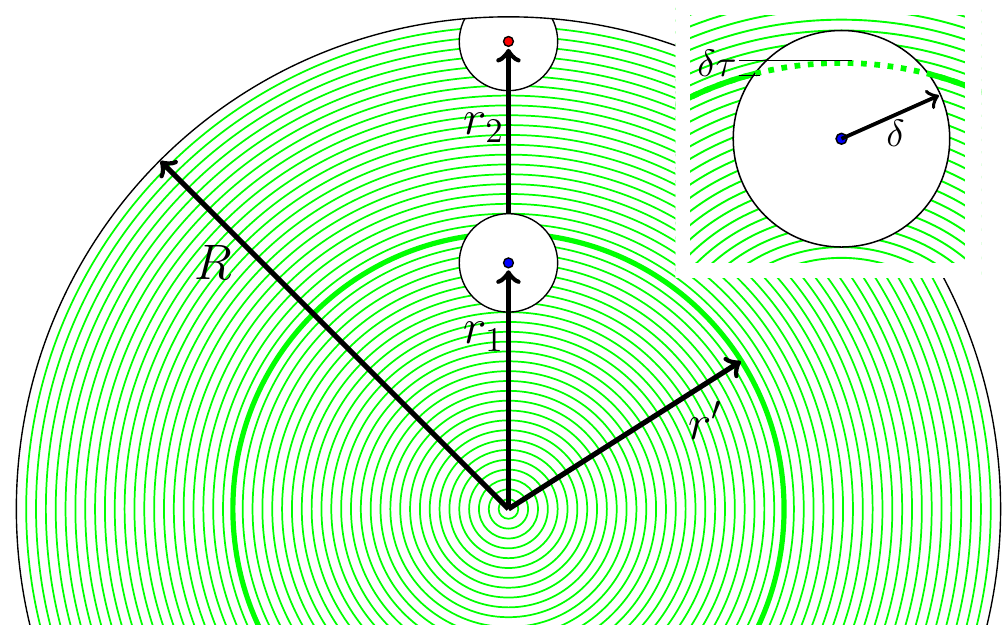}}
 \par\medskip
 \centerline{\includegraphics[width=\columnwidth]{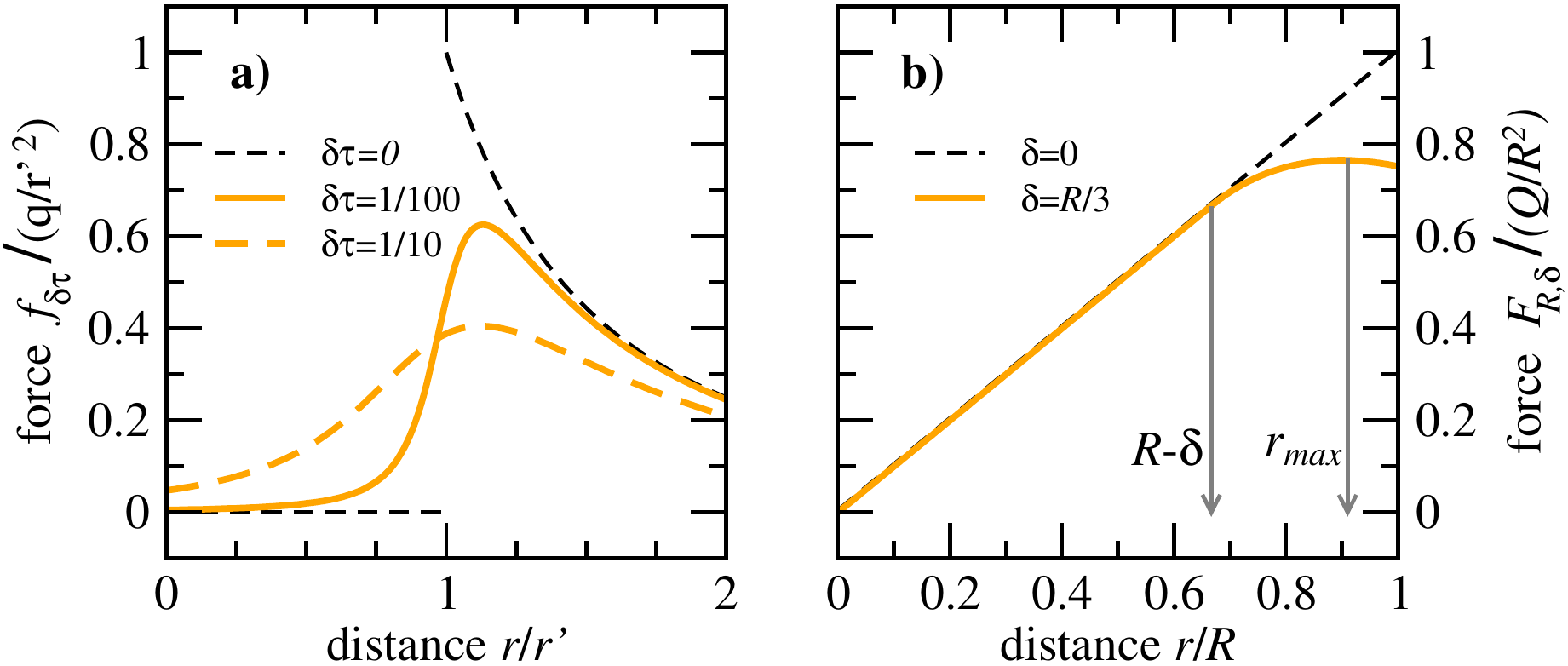}}
 \caption{(Color online) Sketch of the integration assuming a correlation hole with radius $\delta$ around the test particle at distance $\rt$.
 We show two situations, where the correlation hole is inside the bulk ($\rt_{1}$) and at the surface ($\rt_{2}$), respectively.
 The test particle force is obtained by integration over all shells (indicated by green/gray lines) with some of them being ``open'' (thick green/gray line, see also inset) due to the correlation hole.
 The force on a test particle as a function of the distance $\rt$ (a) due to an ``open'' charged spherical shell with radius $r$ according to \eq{eq:frinteg} and (b) integrated over all shells of the charged sphere of radius $R$ according to \eq{eq:chforce}.}
\label{fig:force}
\end{figure}%
Due to the repulsive interaction potential of the atoms in the cluster, nearby particles are very unlikely in the ground state. Hence, a correlation hole around an atom or ion exists.
This means if there is an atom at position $\vrt$ the probability to find another atom at position $\vrs$ within the radius $\delta$ of the correlation hole, i.\,e., for $|\vrt{-}\vrs|\,{<}\,\delta$, vanishes leading to the characteristic pair-correlation function as shown in \fig{fig:spec-100-1000-10000}d.
It has been numerically obtained by an ensemble average over
\begin{equation}\label{eq:mpcf}
g(r)=\frac{R^{3}}{r^{2}\,\tilde{g}(r)}\sum_{i\ne j}\delta(r-r_{ij}),
\end{equation}
with $r_{ij}$ the distance between the particles $i$ and $j$ and 
$\tilde{g}(r)=\frac{3}{16}(r/R{-}2)^{2}(r/R{+}4)$
the distribution of distances $r$ in a sphere with radius $R$ \cite{ga05}.
The normalization with $\tilde{g}(r)$ is necessary in order to remove the trivial dependence on $r$ due to the finite size of the system.

We can assess how the correlation hole affects the force acting on an ion at position $\vrt$ analytically.
To this end we determine the Coulomb repulsion of the ion from an infinitesimally thin spherical shell of radius $\rs$ and charge $q$ by integrating over all angles
\begin{equation}\label{eq:force}
\vf_{\rs}(\vrt)=\frac{q}{4\pi}\int_{0}^{2\pi}\!\!\!\!{\rm d}\phi
\int_{0}^{\pi}\!\!\!{\rm d}\theta\sin\theta\,
\frac{\partial}{\partial\vrt}
\frac{1}{|\vrt-\vrs|},
\end{equation}
with $\vrs\equiv \rs(\sin\theta\cos\phi,\sin\theta\sin\phi,\cos\theta)$ a vector on the shell.
The radial component of this force can be determined without loss of generality by choosing $\vrt = \rt\hat\vz$ along the $z$-axis. Integratig over $\phi$ and using $\tau\equiv\cos\theta$ it reads
\begin{subequations}\mathindent=15pt\label{eq:frinteg}\begin{align}\label{eq:frinteg1}
f_{\rs\!,\delta \tau}(\rt) &= \frac{q}{2}\int_{-1}^{+1-\delta \tau}\!\!\!{\rm d}\tau\; 
\frac{d}{d\rt}\frac{1}{\sqrt{\rt^{2}+\rs^{2}-2\rt\rs \tau}},
\\ \label{eq:frinteg2}
& =\frac{q}{\rt^{2}}
\left[\frac{1}{2}+\frac{(1{-}\delta\tau)\rt-\rs}{2\sqrt{\rs^{2}+\rt^{2}-2(1{-}\delta\tau)\rs\rt}}\right].
\end{align}\end{subequations}
Hereby, the integration was restricted to the upper limit $1{-}\delta \tau\,{<}\,1$ in order to account for a correlation hole, cf.\ the sketch in \fig{fig:force} and its inset, which shows $\delta \tau$ explicitly.  
\Eq{eq:force} corresponds to $\delta \tau = 0$. In this case, performing the integration over $\tau$ yields\,---\,as expected\,---\,Gauss' law, i.\,e.\ $ f_{\rs\!,0} = 0$ for $\rt < \rs$ and $ f_{\rs\!,0} = q/\rt^{2}$ for $\rt\ge\rs$.
As can be seen in \fig{fig:force}a, for finite $\delta\tau$ the force is $f_{\rs\!,\delta\tau}>0$ inside the shell  with radius $\rs$ and $f_{\rs\!,\delta\tau}<q/2\rt^{2}$ outside this shell.
Yet, this modification of the forces does not play a role as long as the test particle's correlation hole is in the bulk ($\rt<R-\delta$, see test particle at $\rt_{1}$ in \fig{fig:force}) since reduced repulsion from inner charged shells with $\rs<\rt$ is fully compensated by a finite repulsion from outer shells with $\rs>\rt$ in accordance with Gauss' law for spherical charge distributions and a test particle with its correlation sphere completely inside the charge distribution. A reduced force is expected at the surface (see test particle at $\rt_{2}$ in \fig{fig:force}). Both expectations are confirmed by the cumulative force, i.\,e., the integration over all charged shells, shown in \fig{fig:force}b.
For this integration we use $\delta\tau = \big(\delta^{2}-(\rt{-}\rs)^{2}\big)/(2\rt\rs)$, which guarantees that the ``open'' shells (see thick green/gray line in the sketch of \fig{fig:force}) form a spherical correlation hole with radius $\delta$ for $|\rt-\rs|\le\delta$. For all other shells it is $\delta\tau=0$. This $\delta\tau$ inserted into \eq{eq:frinteg} and integrated over all shells yields an explicit expression for the force in the presence of a correlation hole
\begin{subequations}\label{eq:chforce}\begin{align}
F_{R,\delta}(\rt) &= Q\rt/R^{3}\phantom{a_{R,\delta}(r)\,}\quad\mbox{for }\rt<R{-}\delta
\label{eq:chforce1}\\
&=A_{R,\delta}(r)\,Q\rt/R^{3}\quad\mbox{for }R{-}\delta<\rt<R
\label{eq:chforce2}
\end{align}
with the dimensionless attenuation factor 
\begin{equation}\label{eq:attentuation}
A_{R,\delta}(r)\equiv
\frac{(\rt{+}R{-}\delta)^{2}\big(2(\rt{+}R)\delta+\delta^{2}-3(\rt{-}R)^{2}\big)}{16\delta\rt^{3}}\,.
\end{equation}
\end{subequations}
Indeed, the force from \eq{eq:chforce} increases linearly, characteristic for a homogeneously charged sphere and without any effect of the correlation hole until the latter touches the surface from the inside (for the case shown in \fig{fig:force}b at $\rt\,{=}\,2R/3$). When this happens the force grows more slowly than in the homogeneous case and even decreases still within the charged sphere reaching a maximal value at $r_{\mathrm{max}}= R-\delta/3$ (for $\delta \ll R$).

With the initial forces sculpturing the properties of the final ion-energy spectrum, we may even attempt to calculate this spectrum according to \eq{eq:spec}.
We assume a self-similar expansion, i.\,e. scaling of all lengths in the system with the common factor $\eta$.
Consequently, the force \eqref{eq:chforce} inherits the property $F_{\eta R,\eta\delta}(\eta\rt) = \eta^{-2}F_{R,\delta}(\rt)$ from the  Coulomb force. Thus, the final kinetic energy depends on the initial position similarly as in the case of a homogenous charge density through an integral along the similarity path with increment ${\rm d}r' = r\,{\rm d}\eta$ 
\begin{equation}\label{eq:emfch}
E(\rt) = r \int_{1}^{\infty}\!\!\!{\rm d}\eta\, F_{\eta R,\eta \delta}(\eta r)=\rt\,F_{R,\delta}(\rt)\,.
\end{equation}
In all (numerical and analytical) spectra presented we take into account a finite energy resolution $\delta E = E_\star/50$. 
For \eq{eq:spec} this means to replace $\delta(x)$ with $K_{\delta E}(x)=\exp(-(x/\delta E)^{2})/\sqrt{\pi}\delta E$.
In \fig{fig:spec-100-1000-10000}b one sees the spectrum for clusters with 1000 atoms in comparison to the one obtained with the mean-field approach including the correlation hole as just described. 
The deviations of both the numerical result (orange/gray-shaded area) and the correlation-hole result  (solid line) with respect to the simple mean-field result (dashed line) from \eq{eq:edcm} at energies $E\lesssim E_{\star}$ clearly reveals the importance of the correlation hole.

\begin{figure}[t!]
\centerline{\includegraphics[width=0.8\columnwidth]{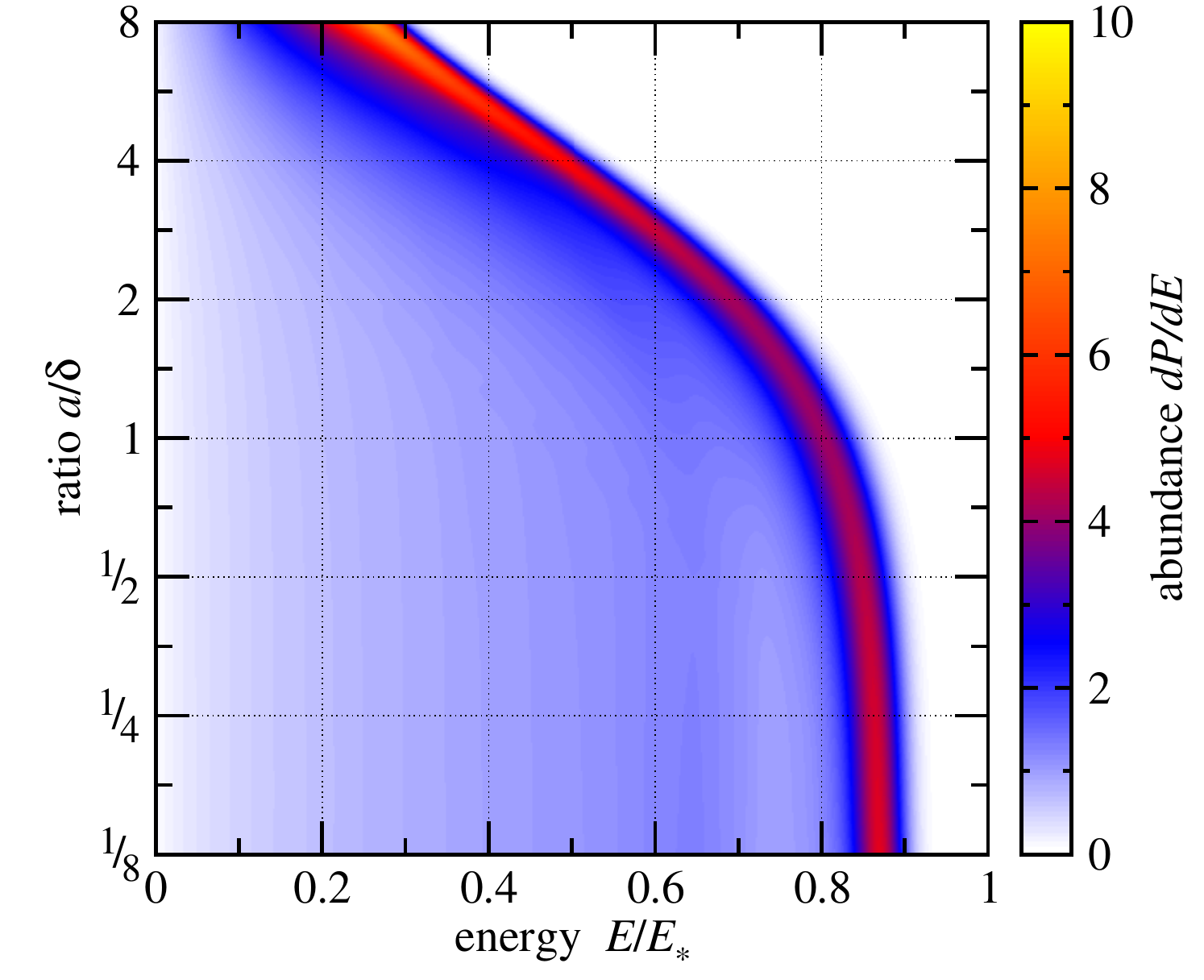}}
 \caption{(Color online) Energy spectra as obtained from an exploding cluster with 1000 particles
 for various cluster edges $a$ (implemented with Fermi distribution function, see text), which is given in units of the correlation hole radius $\delta$. Energy is given in terms of $E_{\star}$, cf.\ \eq{eq:edcm2}. }
\label{fig:spec-lj1000-surface}
\end{figure}%
However, given the analytically calculated force of \eq{eq:chforce2} we also know that
in order for the correlation hole to have an effect, the ion density must have a sharp edge, essentially fall to zero within the radius $\delta$ of the correlation hole. 
Otherwise the forces  of each shell in the presence of the correlation hole will compensate each other, see \fig{fig:force}a. 
On the other hand, a soft edge gives rise to the mean-field peak through catching up of faster inner ions with the slower outer ones, as described in the introduction. We may quantitatively describe the edge by a Fermi distribution $\varrho_{a,R}(r)=2\tilde\varrho_{a,R}/(1+\exp((r{-}R)/a))$ with the softness parameter $a$ and $\tilde\varrho_{a,R}$ the ion density at $r{=}R$ which is determined through the integral $4\pi\int{\rm d}r\,r^{2}\varrho_{a,R}(r){=}N$.  
In \fig{fig:spec-lj1000-surface} the  energy spectrum of ions 
 initially distributed according to a Fermi distribution $\varrho_{a,R}(r)$ is shown for various values of $a$ measured in terms of the correlation hole radius $\delta$.  One can see the cross over from the granularity to the mean-field peak near $a =\delta$. 

We have seen that these two features are of very different origin although both of them are surface effects and occur therefore on the rim of maximal energy in the spectra of \fig{fig:spec-lj1000-surface}.  It should be possible in an experiment to reveal both phenomena and their cross over by simply varying the pulse length of the ionizing light pulse as demonstrated in \fig{fig:spec-lj1000-pulse}. A time-delayed probe pulse could reveal in addition the different temporal behavior of both peaks illustrated in \fig{fig:spec-time}.

In the examples discussed here, the granularity of the system was quantified by the correlation hole which is the most common case for ground state matter. However, a little thought reveals, that ions  (or electrons)
 randomly distributed over a finite volume, give rise to a granularity peak as well. It  is much weaker  and broader \footnote{U. Saalmann, A. Mikaberidze and J.\,M. Rost, unpublished} but still has the same reason: 
 Forces on an ion from other ions inside and outside a virtual shell do not compensate each other near the edge of the sample where the radius of the shell is defined by the distance of the ion to the charge center of the sample.

Acknowledgement\,---\,We gratefully acknowledge support from CORINF, a Marie Curie ITN of the European Union, Grant Agreement No.\,264951.

\end{document}